\documentclass[lettersize,journal]{IEEEtran}

\usepackage{amsmath,amsfonts}
\usepackage{algorithmic}
\usepackage{array}
\usepackage[caption=false,font=normalsize,labelfont=sf,textfont=sf]{subfig}
\usepackage{textcomp}
\usepackage{stfloats}
\usepackage{url}
\usepackage{verbatim}
\usepackage{graphicx}
\def\BibTeX{{\rm B\kern-.05em{\sc i\kern-.025em b}\kern-.08em
    T\kern-.1667em\lower.7ex\hbox{E}\kern-.125emX}}
\usepackage{balance}
\usepackage{orcidlink}
\usepackage{pseudo}
\usepackage{multirow}

\usepackage[normalem]{ulem}

\usepackage{algorithm}
\floatname{algorithm}{Alg.}

\begin{document}

\title{Faster Vertex Cover Algorithms on GPUs with\\Component-Aware Parallel Branching}

\author{
Hussein Amro\textsuperscript{\textdagger},
Basel Fakhri\textsuperscript{\textdagger},
Amer E.~Mouawad\,\orcidlink{0000-0003-2481-4968},
Izzat El Hajj\,\orcidlink{0000-0003-3356-6898}
\thanks{\textsuperscript{\textdagger} The first two authors contributed equally to this work.}
\thanks{All authors are with the Department of Computer Science at the American University of Beirut. Amer E.~Mouawad is also with the David R. Cheriton School of Computer Science at the University of Waterloo. This work was supported by the University Research Board of the American University of Beirut (AUB-URB-104391-26749).}
}

\maketitle

\begin{abstract}

Algorithms for finding minimum or bounded vertex covers in graphs use a branch-and-reduce strategy, which involves exploring a highly imbalanced search tree.
Prior GPU solutions assign different thread blocks to different sub-trees, while using a shared worklist to balance the load.
However, these prior solutions do not scale to large and complex graphs because their unawareness of when the graph splits into components causes them to solve these components redundantly.
Moreover, their high memory footprint limits the number of workers that can execute concurrently.

We propose a novel GPU solution for vertex cover problems that detects when a graph splits into components and branches on the components independently.
Although the need to aggregate the solutions of different components introduces non-tail-recursive branches which interfere with load balancing, we overcome this challenge by delegating the post-processing to the last descendant of each branch.
We also reduce the memory footprint by reducing the graph and inducing a subgraph before exploring the search tree.

Our solution substantially outperforms the state-of-the-art GPU solution, finishing in seconds when the state-of-the-art solution exceeds 6~hours.
To the best of our knowledge, our work is the first to parallelize non-tail-recursive branching patterns on GPUs in a load balanced manner.

\end{abstract}

\section{Introduction}

A vertex cover for a graph is a set of vertices in the graph such that every edge in the graph is incident on at least one of the vertices in the set.
Finding vertex covers of minimum or bounded size has applications in multiple domains such as computational biology, networks, artificial intelligence, and others~\cite{895327,article-vc-app}.
The typical algorithms for finding vertex covers of minimum or bounded size adopt a branch-and-reduce strategy, which ultimately involves exploring a search tree covering the solution space, while pruning the tree using stopping conditions and reduction rules.

Prior works for parallelizing vertex cover algorithms on GPUs~\cite{kabbara2013parallel,abu2018accelerating,yamout2022parallel} assign different branches of the search tree to different thread blocks, which explore the branches concurrently using private stack data structures.
Load balance is particularly challenging because of the highly imbalanced nature of the search tree, so the state-of-the-art GPU solution~\cite{yamout2022parallel} uses a shared worklist to offload tree nodes from busy thread blocks to idle thread blocks.
Still, even the state-of-the-art GPU solution struggles to scale to large and complex graphs due to two limitations.
The first limitation is that it is unaware of when the graph splits into multiple components during the search, causing it to solve the same component redundantly.
The second limitation is that its high memory footprint prevents it from launching enough workers to fully utilize the GPU.

We propose a novel GPU solution for vertex cover problems that overcomes these two limitations.
To overcome the first limitation, we detect when a graph splits into multiple components during the search and branch on different components independently to avoid processing them redundantly.
The main challenge is that the need to aggregate the solutions of different components introduces non-tail-recursive branches into our search tree.
Non-tail recursive branches interfere with the ability to send different children to different thread blocks for the purpose of load balancing.
To address this challenge, we register every non-tail-recursive branch on components in a shared space and track the active descendants of a branch, then delegate the post-processing to the thread block that solves the last descendant.
Our solution works for multiple nesting of non-tail-recursive branches.

To overcome the second limitation, we aggressively apply reduction rules to the graph and induce a subgraph prior to exploring the search tree.
Using the induced subgraph allows us to decrease the memory footprint of the intermediate graph representation, thereby unleashing more parallelism when scaling to large graphs.
We also leverage sparsity in the intermediate graph representation to reduce the number of vertices we visit, and we use small integer data types where possible to further reduce the memory footprint.

Our evaluation shows that our GPU solution for vertex cover substantially outperforms the state-of-the-art GPU solution, finishing in seconds or minutes when the state-of-the-art solution times out after 6~hours.
It demonstrates that our component-aware branching significantly reduces the number of search tree nodes visited, and that our optimizations to the intermediate graph representation enable many more thread blocks to participate in the computation.

Beyond vertex cover algorithms, numerous works parallelize branching algorithms on GPUs~\cite{chen2020pangolin,chen2021sandslash,kawtikwar2023beep,jenkins2011lessons,wei2021accelerating,almasri2023parallelizing,almasri2022parallel} and load balancing is a common challenge that they deal with.
However, all these works parallelize branching patterns that are tail recursive.
To the best of our knowledge, our work is the first to parallelize non-tail-recursive branching patterns on GPUs, while retaining load balancing capabilities.
Our code has been open sourced to enable reproducibility and further research on parallel branching algorithms for GPUs\footnote{\url{https://github.com/HusseinAmro/component-aware-vertex-cover-gpu}}.

\section{Background}\label{sec:background}

\subsection{Definitions}

Let $G = (V, E)$ denote an undirected simple graph with vertex set $V(G)$ and edge set $E(G)$. Let $N(v)$ denote the set of vertices adjacent to a vertex $v \in V(G)$, also called the \emph{neighborhood} of $v$. Let 
$d(v)$ denote the number of edges incident on a vertex $v \in V(G)$, also called the \emph{degree} of $v$.
$\Delta(G)$ denotes the maximum degree in $G$, i.e., the degree of a vertex with the highest degree.
Given a set of vertices $S \subseteq V(G)$, $G - S$ denotes the subgraph induced on the set of vertices $V(G) \setminus S$.

A \emph{vertex cover} for a graph $G$ is a set of vertices $S \subseteq V(G)$, such that for every edge $v_1 v_2 \in E(G)$, $S$ contains at least one of $v_1$ or $v_2$, i.e., $\{v_1, v_2\} \cap S \neq \emptyset$.
The \emph{Minimum Vertex Cover (MVC)} problem is to find a vertex cover $S$ for $G$ such that there is no vertex cover $T$ for $G$ where $|S| > |T|$.
The \textit{Parameterized Vertex Cover (PVC)} problem is to find, for a given integer $k > 0$, a vertex cover $S$ for $G$ such that $|S| \leq k$ (if such a vertex cover exists).

\subsection{Vertex Cover Algorithms}\label{sec:background-alg}

The typical algorithms for solving the MVC and PVC problems adopt the \textit{branch-and-reduce} strategy.
Branch-and-reduce algorithms explore the solution space of a problem by successively breaking the problem into smaller alternative sub-problems (i.e., branching), and solving these sub-problems recursively.
For vertex cover problems, we observe that for any vertex in the graph, a vertex cover must contain either that vertex or all of the vertex's neighbors to ensure that all the edges incident on that vertex are covered.
This observation is the basis for breaking a problem into two alternative sub-problems when branching: one where the solution contains a vertex of our choice and one where the solution contains all of that vertex's neighbors.
A convenient vertex to choose for branching is a vertex of maximum degree, which ensures the largest number of edges is covered when we branch.

\begin{figure}
    \centering
    \includegraphics[width=\columnwidth]{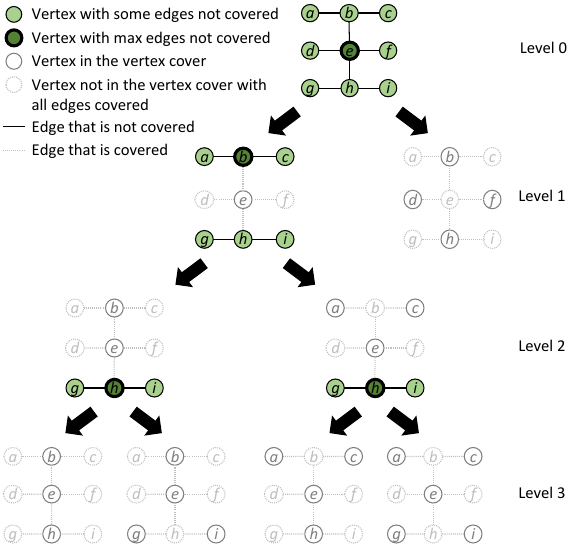}
    \caption{Example of the vertex cover search tree}\label{fig:mvc-example}
\end{figure}

The branching pattern in these algorithms forms a search tree, where every internal node in the search tree is an intermediate sub-problem, and the leaf nodes are solutions or base cases of the recursion.
Figure~\ref{fig:mvc-example} shows one possible search tree for finding the minimum vertex cover of the example graph at the top of the figure (Level~0).
First, we select a vertex of maximum degree, vertex $e$, as the basis for branching.
The left branch includes $e$ in the vertex cover, while the right branch includes $N(e)$ in the vertex cover.
The right branch already finds a solution with the four vertices $\{b, d, f, h\}$ that cover all the edges.
On the other hand, the left branch still has edges that are not covered and proceeds.
We branch on vertex $b$ this time, including $b$ in the solution on the left branch, and $N(b)$ in the solution on the right branch.
We keep repeating this process until all edges in the graph are covered.
Ultimately, the smallest vertex cover we find is the leftmost branch in Level~3 containing the three vertices $\{b, e, h\}$.

Since the number of candidate solutions can increase exponentially with the size of the graph, various strategies can be followed to prune the search tree.
For example, if we find the leftmost branch in Level~3 before processing the rightmost branch of Level~2, then we know that we do not need to continue branching from the latter.
The reason is that the latter already includes three vertices in the vertex cover, so branching can only increase that number.
However, we already know that a vertex cover of size three exists from the leftmost branch of Level~3, so it is futile to search for vertex covers that are larger in size.

Another way to reduce the size of the search tree is by applying \textit{reduction rules}.
These rules quickly identify vertices that must be in the vertex cover to avoid branching on those vertices.
For example, the \textit{degree-one} reduction rule observes that a degree-one vertex only covers one edge, whereas its neighbor covers that same edge and potentially other edges.
Hence, it is always better to include the neighbor in the vertex cover.
The \textit{degree-two triangle} rule observes that we must include at least two vertices in a triangle to cover all its edges.
If a degree-two vertex was part of a triangle, then it is always better to include its two neighbors than to include it since its neighbors may cover potentially other edges.
The \textit{high-degree} rule observes that if the degree of a vertex is higher than the number of vertices needed to exceed the current best solution, then it is always better to include the high degree vertex in the solution.
Including its neighbors is futile since it will not yield a solution better than the one we have already found.

\begin{algorithm}[t]
  \small
  \caption{Algorithm for the Minimum Vertex Cover problem~\cite{yamout2022parallel}}  
  \label{alg:mvc-code}
  \begin{pseudo}
\textbf{function} $\textsf{MVC}(G, S, best)$ \\+
    $(G, S) = reduce(G, S, best)$ \\
    \textbf{if} $|S| \geq best \vee |E(G)| > (best - |S| - 1)^2$ \\+
        \textbf{return} // No MVC on this branch (do nothing) \\-
    \textbf{else if} $|E(G)| == 0$ // New MVC found \\+
        $best = |S|$ \\
        \textbf{return} \\-
    \textbf{else} // Vertex cover not found, need to branch \\+
        \textbf{Let} $v_{max} \in \{u \in V(G) \mid d(u) = \Delta(G)$\} \\
        $\textsf{MVC}(G - \{v_{max}\}, S \cup \{v_{max}\}, best)$ \\
        $\textsf{MVC}(G - N(v_{max}), S \cup N(v_{max}), best)$\\
        \textbf{return}
\end{pseudo}
\end{algorithm}

Algorithm~\ref{alg:mvc-code} shows the pseudocode for a branch-and-reduce algorithm for solving the MVC problem.
$G$ denotes the graph being solved, $S$ denotes the vertex cover solution under construction, and $best$ is the size of the smallest vertex cover found so far (line 1).
The graph $G$ and set $S$ are passed by value, whereas the variable $best$ is passed by reference so that all invocations update the same value.
At the initial invocation, $G$ is the whole graph, $S$ is $\emptyset$, and $best$ is an approximate minimum computed by an approximate algorithm such as a greedy one.
The first step is to apply reduction rules on the graph (line 2).
The vertices added to the solution by the reduction rules are removed from the graph along with their incident edges.
Next, we stop branching if a stopping condition~\cite{yamout2022parallel} is met that deems it infeasible to find a minimum vertex cover on this branch (lines 3-4).
Otherwise, if the graph has no more edges, then we have found a new minimum vertex cover so we update the best solution found so far (lines 5-7).
Otherwise, if we have not yet found a vertex cover, then we need to branch (lines 9-12).
To branch, we select a vertex $v_{max}$ with the highest degree (line 9), and we recursively call MVC twice: once on a sub-problem that includes the vertex $v_{max}$ in the solution (line 10) and once on a sub-problem that includes $N(v_{max})$ in the solution (line 11).

The stopping condition on line 3 of Algorithm~\ref{alg:mvc-code} checks for one of two conditions.
The first condition is if the solution under construction is already as large as the best solution found so far ($|S| \ge best$).
The second condition is if the number of vertices that can still be removed before reaching the current best solution size ($|S| - best - 1$) is sufficient to cover the total number of remaining edges.
After applying the high-degree rule, each vertex cannot have more than $|S| - best - 1$ remaining edges, so the maximum number of edges that can be covered by $|S| - best - 1$ vertices is $(|S| - best - 1)^2$.
If the number of remaining edges ($|E(G)|$) exceeds this number, then no solution can be found on this branch.

The pseudocode for solving the PVC problem is very similar, with two key differences.
First, the reduction rules and stopping conditions are based on $k$, not $best$, since we are looking for a vertex cover of size at most $k$, not a vertex cover smaller then $best$.
Second, when we find a vertex cover of size at most $k$, we end the search, unlike MVC which continues the search to ensure that no smaller vertex cover exists. 

\subsection{Prior GPU Solutions}\label{sec:background-prior}

Prior works for parallelizing MVC and PVC on GPUs~\cite{kabbara2013parallel,abu2018accelerating,yamout2022parallel} rely on having different thread blocks explore different branches of the search tree concurrently, with each thread block using its own private stack.
Earlier approaches~\cite{abu2018accelerating,kabbara2013parallel} extract sub-trees at a specific depth of the search tree and assign different sub-trees to different thread blocks.
However, since the MVC search tree tends to be highly imbalanced, these approaches suffer from high load imbalance because different thread blocks are assigned to explore different sub-trees that are substantially different in size.
The state-of-the-art solution by Yamout et al.~\cite{yamout2022parallel} improves on these earlier approaches by introducing a worklist-based load balancing scheme.
Busy thread blocks add search tree nodes to a shared multi-producer multi-consumer queue~\cite{kerbl2018broker} instead of their private stack, and idle thread blocks pick up these search tree nodes and explore them.

While the load balancing technique in the state-of-the-art GPU solution substantially improves performance, the overall implementation still struggles to scale to large and complex graphs.
We identify two major limitations in the state-of-the-art GPU solution that inhibit its scalability.
The first limitation is that prior works are unaware of when graphs split into multiple components, causing them to solve the same component redundantly in multiple branches.
The second limitation is that as graphs get larger, more memory is needed per thread block to explore a sub-tree, which limits the number of thread blocks that can execute concurrently as well as the ability to keep the intermediate representation of the graph in shared memory.
We describe each of these limitations in more detail and propose solutions to address each limitation in Sections~\ref{sec:components} and~\ref{sec:memory}.

\section{Component-Aware Search Tree Exploration}\label{sec:components}

\subsection{Motivation}

As the vertex cover search tree is being explored, the intermediate graphs tend to split into multiple components.
When a graph consists of multiple components, its minimum vertex cover is naturally the combination of the minimum vertex covers of each component.
Prior GPU solutions~\cite{kabbara2013parallel,abu2018accelerating,yamout2022parallel} continue to treat the graph as a whole graph.
However, it is more efficient to solve each component separately and combine the result than it is to continue treating the graph as one whole~\cite{alsahafy2019computing,hespe2021targeted}.

For example, in the search tree in Figure~\ref{fig:mvc-example}, in the leftmost node of Level~1, the edges that are not covered induce a graph that contains two components: $\{a, b, c\}$ and $\{g, h, i\}$.
Branching on vertex $b$ eliminates the component $\{a, b, c\}$ in both branches, leaving just the component $\{g, h, i\}$ in both nodes at Level~2.
Both nodes at Level~2 subsequently branch on vertex $h$ to eliminate the component $\{g, h, i\}$.
Here, we observe that component $\{g, h, i\}$ was solved twice redundantly.
If we had more components, then this redundancy grows exponentially with the number of components.

\begin{figure}
    \centering
    \includegraphics[width=\columnwidth]{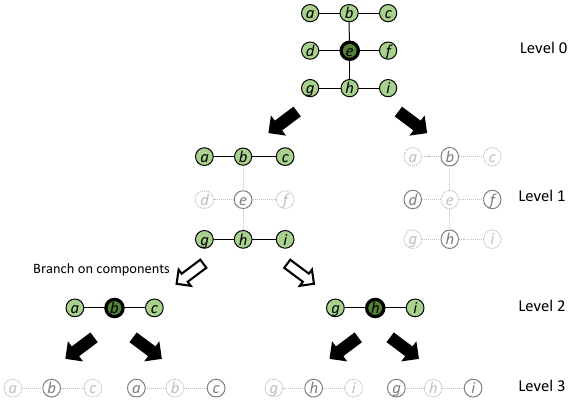}
    \caption{Example of vertex cover search tree with component-aware exploration}\label{fig:mvc-example-component}
\end{figure}

To avoid such redundancy, the search tree can branch on each component separately then combine the result of each component.
Such a component-aware search tree exploration is shown in Figure~\ref{fig:mvc-example-component}.
Levels~0 and~1 are the same as Figure~\ref{fig:mvc-example}.
However, when the leftmost node at Level~1 realizes that the graph has split into two components, it creates an independent branch for each component.
Hence, Level~2 of the search tree has a node for component $\{a, b, c\}$ and a node for component $\{g, h, i\}$.
Accordingly, the component $\{g, h, i\}$ is solved only once instead of twice.
The examples in Figure~\ref{fig:mvc-example} and Figure~\ref{fig:mvc-example-component} have the same number of tree nodes because the redundant branch on component $\{g, h, i\}$ that we save from Figure~\ref{fig:mvc-example} is offset by the additional branch on components in Figure~\ref{fig:mvc-example-component}.
However, this similarity is only an artifact of the example being small for illustration purposes.
If component $\{g, h, i\}$ required a large number of branches to be solved, these branches would occur twice in Figure~\ref{fig:mvc-example} but only once in Figure~\ref{fig:mvc-example-component}, leading to a substantial reduction in tree nodes.

From a theoretical standpoint, the worst-case time complexity of branching without component-awareness is $\mathcal{O}^*(\beta^n)$ where $\beta$ is the effective branching factor ($1 < \beta < 2$) and $n$ is the number of vertices.
While the time complexity remains exponential in $n$ when component-awareness is introduced, it lowers the effective branching factor $\beta$.
At split nodes, assuming (pessimistically) that each split yields at most two components, rather than solving a size $n$ instance we solve two sub-problems of sizes $n_1+n_2=n$, so the local work transforms from $\mathcal{O}^*(\beta^n)$ to $\mathcal{O}^*(\beta^{n_1}+\beta^{n_2})$, which is strictly smaller since $0<n_1,n_2<n$ and $\beta>1$.
Aggregated over the tree, these splits lower the effective branching factor from $\beta$ to $\beta_{e} \approx \beta^{\,1-\rho\,\eta}$, where $\rho$ is the fraction of internal nodes where splits occur, and $\eta\in(0,1]$ reflects how balanced the two components are (larger when $n_1\approx n_2$).\footnote{Our modeling of the effective branching factor reflects standard analyses in exact algorithms where component splits reduce exponential growth. Fomin et al.~\cite{fomin2009measure} show that decomposing graphs into connected components can lower the time bound for computing vertex covers from $O(2^{0.465n})$ to $O(2^{0.398n})$ by reducing the size of subproblems. Similarly, Niedermeier~\cite{niedermeier2006invitation} explains how such splits lead to additive recursions that slow the growth of the search tree. Our model captures this effect by quantifying the frequency $\rho$ and balance $\eta$ of splits, consistent with these analyses.}
For example, with baseline $\beta=1.50$, very rare splits $\rho=0.02$, and moderately imbalanced partitions ($\eta=0.5$), the effective branching factor $\beta_{e}=1.5^{0.99} \approx 1.494$.
For a small number of vertices $n=200$, this lowered effective branching factor yields about $(\beta/\beta_{e})^{200} \approx 2.25\times$ fewer search tree nodes.
In practice, the split rate $\rho$ can be higher, splits can yield more than two components, and the number of vertices $n$ is much larger, which results in dramatic reductions in the number of search tree nodes visited as we show in Table~\ref{tab:component} of Section~\ref{sec:eval-components}.

The pseudocode for component-aware exploration is shown in Algorithm~\ref{alg:mvc-code-component}.
The first eight lines resemble the pseudocode in Algorithm~\ref{alg:mvc-code}.
The difference is on the subsequent lines which contain the branching strategy.
When we need to branch, the pseudocode for component-aware exploration first finds the components (line 9).
If the graph has only one component (line 10), then it proceeds like the code in Algorithm~\ref{alg:mvc-code} by branching on a vertex of maximum degree (lines 11-13).
On the other hand, if the graph has multiple components (line 14), then the code branches on each component separately (lines 15-20).

While branching on components, we need to track the total size of the solution as each component contributes to it.
We do so with a variable $sum$ that is initialized to the number of vertices in the solution so far before branching (line 15).
We then iterate over the components (line 16).
For each component $i$, we allocate and initialize a variable $best_i$ to track the size of a minimum vertex cover of that component (line 17).
The first bound on $best_i$ is the number of vertices remaining that can be added to the solution before exceeding the current best ($best - sum$).
The second bound is one less than the total number of vertices in the component.
We then call MVC recursively on the component (line 18).
After finding a minimum vertex cover of the component, we add its size to $sum$ (line 19).
After processing all components, if the size of the new vertex cover found $sum$ is less than the best solution so far, we set it as the new best solution (line 20).

\begin{algorithm}[t]
    \small 
    \begin{pseudo}
\textbf{function} $\textsf{MVC}(G, S, best)$ \\+
    $(G, S) = reduce(G, S)$ \\
    \textbf{if} $|S| \geq best \vee |E(G)| > (best - |S| - 1)^2$ \\+
        \textbf{return} // No MVC on this branch (do nothing) \\-
    \textbf{else if} $|E(G)| == 0$ // New MVC found \\+
        $best = |S|$\\
        \textbf{return}\\-
    \textbf{else} // Vertex cover not found, need to branch \\+
        $C = findComponents(G)$\\
        \textbf{if} $|C| == 1$ // Graph has one component\\+
            \textbf{Let} $v_{max} \in \{u \in V(G) \mid d(u) = \Delta(G)$\} \\
            $\textsf{MVC}(G - v_{max}, S \cup \{v_{max}\}, best)$ \\
            $\textsf{MVC}(G - N(v_{max}), S \cup N(v_{max}), best)$\\-
        \textbf{else} // Graph has multiple components\\+
            $sum = |S|$\\
            \textbf{for} $G_i \in C$\\+
                $best_i = min(best - sum, |V(G_i)| - 1)$\\
                $\textsf{MVC}(G_i,\emptyset, best_i)$\\
                $sum = sum + best_i$\\-
                
            $best = min(sum, best)$ \\
\end{pseudo}
    \caption{Algorithm for {\sc Minimum Vertex Cover} with component-aware exploration}\label{alg:mvc-code-component}
\end{algorithm}

In the rest of this section, we propose a parallel GPU implementation of component-aware exploration.
Our implementation is built on top of the state-of-the-art solution~\cite{yamout2022parallel} with three major changes.
The first change introduces a routine for identifying when a graph has broken into multiple components and finding those components (Section~\ref{sec:components-finding}).
The second change modifies the branching mechanism to branch on components while retaining load balancing capabilities (Section~\ref{sec:components-branching}).
The third change adds reduction rules that target special types of components (Section~\ref{sec:components-rules}).
We also discuss the changes needed to support PVC (Section~\ref{sec:components-pvc}).

\subsection{Finding Components}\label{sec:components-finding}

For a thread block to find the components of a graph that it is processing (line 9 in Algorithm~\ref{alg:mvc-code-component}), we use multiple parallel breadth-first-search executions.
We first identify a vertex in the graph whose degree is not zero and set it as the source of the breadth-first search.
The threads in the block collaboratively execute a vertex-centric pull-based breadth first search~\cite{wen2022programming} and mark all visited vertices which form a component.
If all vertices are visited, we conclude that we have one component and proceed normally by branching on a vertex of maximum degree.
On the other hand, if not all the vertices with non-zero degree are visited, we conclude that we have multiple components.
If we have multiple components, we initiate a branch on components (see Section~\ref{sec:components-branching}) and send the component that we just found to the thread block's private stack or to the shared worklist for processing.
We then repeatedly execute breadth-first-search to find components and send them for processing until all vertices have been visited, which indicates that all components have been found.
Importantly, we do not find all components and then send them together to be processed, but we send the components eagerly as soon as we find them.
Sending components for processing as soon as they are found ensures that components can be solved in parallel while new components are still being found.
We also tried performing a parallel connected components operation instead of a parallel breadth-first search, but the breadth-first search approach performed better because it found components more eagerly.
\looseness=-1

\subsection{Branching on Components}\label{sec:components-branching}

The branching pattern of component-aware exploration is challenging to support under prior work's~\cite{yamout2022parallel} load balancing scheme.
In prior work, when a node in the search tree is processed and needs to branch, child nodes can be offloaded to other thread blocks to achieve load balancing.
This offloading relies on the fact that the children nodes do not need to report any result back to the parent node for post-processing, so the children can be completely disowned to other thread blocks.
From the perspective of the pseudocode in Algorithm~\ref{alg:mvc-code}, the recursive calls to MVC are tail-recursive and there is no post-processing needed when the calls return.

In contrast, the component-aware exploration in Algorithm~\ref{alg:mvc-code-component} does not have this tail-recursive property.
After the recursive calls return, the parent needs to post-process the results of the children nodes to contribute their results to a new minimum vertex cover (lines 19-20).
One solution could be to forgo load-balancing and keep the parent and all its descendants in the same thread block.
In this case, the thread block would send all the parent's components to its private stack, process them one after the other, then do the post-processing after they have all completed.
However, we show in Section~\ref{sec:eval-lb} that the performance of such a scheme suffers due to the loss of load balance.

To overcome this challenge, we propose to perform the post-processing activities in the children themselves.
One aspect of the post-processing is accumulating the size of a component's minimum vertex cover to the parent's sum (line 19 in Algorithm~\ref{alg:mvc-code-component}).
This operation must be done for each component, so we have the last descendant for each component do it for that component.
Another aspect of the post-processing is updating the best solution if we identify that the sum of all the components' minimum vertex covers leads to a better solution (line 20 in Algorithm~\ref{alg:mvc-code-component}).
This operation must be done once after processing all the components, so we have the last descendant of the last component do it on behalf of the parent.
These techniques are analogous to converting the non-tail-recursive calls into tail-recursive calls by moving the post-processing after the calls into the bodies of those calls.

To be able to achieve the aforementioned techniques, we need a mechanism to identify the last descendant to execute for each component, as well as the last component to be solved.
Moreover, we need a mechanism for these descendants to find their components' $best_i$ as well as their parent's $sum$ variable to be able to update them.
We provide such a mechanism by introducing a \textit{component branch registry}.
The component branch registry is a list of metadata entries in global memory that include information needed about branches on components.
Whenever we branch on components, we register the branch by adding entries to the registry for the parent node as well as for each of the children nodes (i.e., components) involved in the branch.
The children nodes pass a reference to their entries to all their descendants.

Each entry in the component branch registry consists of three integers.
For the children nodes, the three integers are:
\begin{itemize}
    \item \textit{Best}: The size of the best solution found so far for this component (i.e, $best_i$ in Algorithm~\ref{alg:mvc-code-component})
    \item \textit{LiveNodes}: A counter for the number of descendants of the child node that are still executing
    \item \textit{ParentIdx}: The index of the parent node's entry in the registry
\end{itemize}
As the child node and its descendant solve the component, they update \textit{Best} as the solution for that component.
Moreover, every time the child node or one of its descendants branches on a vertex, it increments \textit{LiveNodes}, and every time a node completes, it decrements \textit{LiveNodes}.
All these updates are performed atomically.
When \textit{LiveNodes} reaches 0, the node that decremented it knows that it is the last descendant solving the component and is responsible for doing the post-processing.
Accordingly, it uses \textit{ParentIdx} to find the entry of the parent node of the component branch.

For the parent node, the three integers in the component branch registry entry are:
\begin{itemize}
    \item \textit{Sum}: The number of vertices added to the solution by the parent and all its components so far (i.e, $sum$ in Algorithm~\ref{alg:mvc-code-component})
    \item \textit{LiveComps}: A counter for the number of components of the parent node that are still being solved.
    \item \textit{AncestorIdx}: The index of the parent node's ancestor's entry in the registry, since the parent node itself may be a descendant of an earlier branch on components
\end{itemize}
When the last descendant of a child node obtains the solution for a component, it adds the size of that solution to \textit{Sum} (i.e., it performs line 19 in Algorithm~\ref{alg:mvc-code-component}).
It also decrements \textit{LiveComps} to indicate that one of the parent's components has been solved.
All these updates are performed atomically.
When \textit{LiveComps} reaches 0, the node that decremented it knows that it has solved the last component of the parent and is responsible for doing further post-processing.
Accordingly, it uses \textit{AncestorIdx} to find the entry of the parent node's ancestor (which itself may be a child of another branch) and updates that ancestor's $Best$ value (i.e., it performs line 20 in Algorithm~\ref{alg:mvc-code-component}) as well as decrements the ancestor's $LiveNodes$ value.
This process can be iterative since $LiveNodes$ for the ancestor could reach zero, triggering another sequence of updates.
In this way, we support multiple nesting of branches on components that emerge within larger components.

\begin{figure}
    \centering
    \includegraphics[width=\columnwidth]{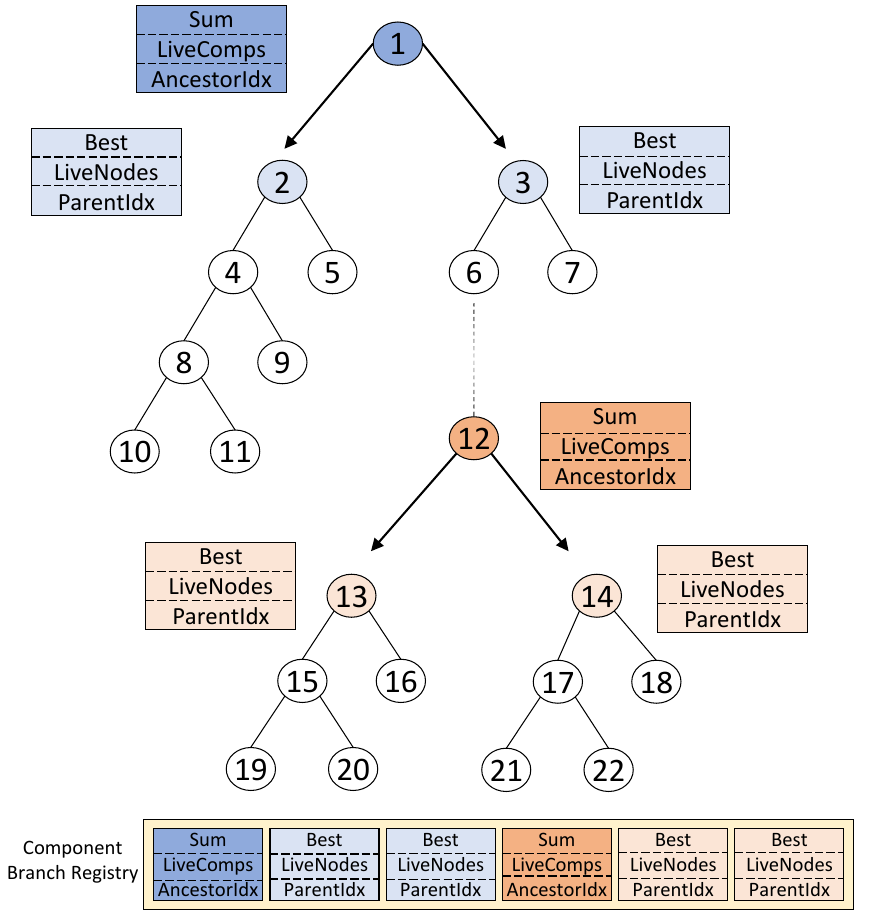}
    \caption{Example of nested branching on components using the component branch registry}\label{fig:registry-example}
\end{figure}

Figure~\ref{fig:registry-example} shows an example of how we branch on components using the component branch registry.
In this example, tree node~1 finds two components in its graph, so it creates two child nodes~2 and ~3, one for each component.
A parent entry is created in the registry for node~1, and child entries are created for each of nodes~2 and~3.
The branches rooted and nodes~2 and~3 each search for a solution to their component.
The descendants of nodes~2 and~3 all access the $Best$ and $LiveNodes$ variables in the nodes' registry entries to track the best solution found for the component and the number of active descendants, respectively.
Eventually, one of the descendants of node~3, namely node~12, finds two components.
It then creates a branch for each component, namely nodes~13 and~14.
Entries are then created in the registry for nodes~12, 13, and~14.
The descendants of nodes~13 and~14 access the two nodes' registry entries while searching for a solution.

Now let's assume that node~20 is the last descendant of node~13 to be visited.
In this case, when the thread block that completes node~20 goes to node~13's registry entry and decrements $LiveNodes$, it observes that $LiveNodes$ becomes zero and realizes that it just processed the last descendant.
Accordingly, the thread block uses node~13's $ParentIdx$ to find the registry entry for node~12.
It adds node~13's $Best$ value to node~12's $Sum$ value, and decrements node~12's $LiveComps$ value to indicate that the component assigned to node~12 has been solved.
If node~14's component is still being solved, the updated value of $LiveComps$ will be one so the thread block is done and can grab other work from its stack or the worklist.
On the other hand, if node~14's component has already been solved, then the updated value of node~12's $LiveComps$ will be zero, indicating that the thread block has just finished solving the last component of node~12.
In this case, the thread block is responsible for checking if the sum across components has yielded a new best solution.

To do so, the thread block uses node~12's $AncestorIdx$ to find the entry for the component that it is solving, namely node~3's entry.
If node~12's $Sum$ is better than node~3's $Best$, then the thread block updates node~3's $Best$ to become node~12's $Sum$.
The thread block also decrements node~3's $LiveNodes$ to indicate that node~12 is complete.
Here, it is possible that node~12 is the last descendant of node~3, causing node~3's $LiveNodes$ to become zero.
In this case, the thread block repeats the process, contributing node~3's $Best$ to node~1's $Sum$, decrementing node~1's $LiveComps$, and even propagating node~1's $Sum$ further upward if node~1's $LiveComps$ has reached zero.

Importantly, as mentioned in Section~\ref{sec:components-finding}, some components are solved while other components are still being found.
To avoid a situation where all the components found so far are solved and \textit{LiveComps} becomes zero before all the remaining components are found, we count the parent node that is searching for components among the \textit{LiveComps}, and we decrement \textit{LiveComps} when the parent node finishes searching for components.
Accordingly, it is impossible for \textit{LiveComps} to reach zero if the parent node is still active and searching for components.

With these mechanisms in place, the child nodes of a branch on components and their descendants can be solved by arbitrary thread blocks regardless of where the parent node was solved.
Accordingly, a thread block that solves the parent node of a branch on components can send its child nodes to the worklist for load balancing, rather than keep them in its private stack to solve them by itself.

\subsection{Component-Targeting Reduction Rules}\label{sec:components-rules}

When the graph breaks into multiple components, some of the small and simple components can be solved entirely by the reduction rules.
Prior work applies various reduction rules including the degree-one rule, degree-two triangle rule, and high-degree rule.
However, we have identified certain types of components that cannot be handled by these reduction rules, namely cliques and chordless cycles.
For this reason, we introduce two new reduction rules to handle each of these types of components.

A clique is a subgraph where every vertex in the clique is connected to every other vertex with an edge.
We can easily identify if a component is a clique by checking if all the vertices in the component have a degree equal to the number of vertices in the component minus one.
If a component is a clique, then we simply add all but one vertex of the clique to the solution to cover all the clique's edges.

A chordless cycle is a cycle where no two vertices in the cycle are connected by an edge that is not part of the cycle.
We can easily identify if a component is a chordless cycle by checking if all the vertices in the component have a degree of two.
If a component is an isolated chordless cycle, then we must add half the vertices in the cycle to the solution to cover all the cycle's edges (we take the ceiling if the number of vertices is odd).

\subsection{Supporting the Parameterized Variant}\label{sec:components-pvc}

Recall from Section~\ref{sec:components-branching} that when a node solving a component finds a smaller vertex cover than the best solution so far, it updates \textit{Best} in the component's entry in the component branch registry and proceeds with searching for smaller solutions to the component.
Only the last descendant contributes this update to the parent.
The reason is that with MVC, we need to exhaust the search to make sure that we find a smallest possible solution.

In contrast, in PVC, we do not need to exhaust the search.
We can stop as soon as we find a vertex cover that satisfies the parameter $k$.
However, the parameter $k$ is a limit for the entire solution, and we cannot partition it across components.
For this reason, in PVC, whenever a node solving a component finds a smaller solution, it does not only update its own component's \textit{Best} value.
It follows the chain of registry entries to propagate this update all the way up to the root node.
If the new solution at the root node reaches $k$, then the search can terminate, otherwise we continue exploring the component.
Hence, compared to MVC, PVC trades off more frequent update propagations to the root node for the sake of terminating the search early.

Our implementation of PVC is for the standard problem of finding one vertex cover of size at most $k$.
An alternative formulation is to enumerate all vertex covers of size at most $k$.
We currently do not support this variant, and neither do any of the prior GPU implementations~\cite{kabbara2013parallel,abu2018accelerating,yamout2022parallel}.
Supporting this variant would require enumerating the vertex covers of each component whose size is at most $k$ (a tighter bound is possible), then generating all combinations of vertex covers whose size is at most $k$.
It would require changing the stopping condition, allocating additional memory to store the enumerated variants, and extending the post-processing step to generate the combinations of variants.

\section{Optimizing the Degree Array Representation}\label{sec:memory}

\subsection{Motivation}

As the vertex cover search tree is explored, the intermediate state of the graph needs to be retained at each pending node in the search tree.
This intermediate state tracks which vertices have been removed from the graph and added to the solution, and which vertices remain and are candidates for removal.
In prior works, the graph is represented using the Compressed Sparse Row (CSR) format, and the intermediate state is represented as a \textit{degree array}, where an integer is stored for each vertex in the graph representing the vertex's degree.
\looseness=-1

As graphs get larger, the size of the degree array representing an intermediate graph also gets larger.
This increase in the size of the degree array has consequences that degrade performance.
The first consequence is that it limits the number of thread blocks that can run concurrently.
As thread blocks explore their sub-trees in a depth-first manner, each thread block maintains a local stack of the degree arrays of the pending nodes in the exploration.
Larger degree arrays result in fewer stacks that can be stored simultaneously, and hence, fewer thread blocks that can execute concurrently.
The second consequence is that it prevents the use of shared memory for storing the intermediate graph of a tree node being solved.
When a thread block removes a degree array from its private stack or the shared worklist for processing, it places that degree array in shared memory for fast access.
However, if the degree array is too large, the thread block falls back on placing it in global memory which results in slower accesses.
The third consequence is that it increases the number of unnecessary degree checks performed.
While the degree array representation is compact, quick to access, and easy to maintain, it does not capture sparsity, which means that zero-degree vertices are checked every time when applying the reduction rules.
Towards the bottom of the search tree, most vertices have a degree of zero and are checked unnecessarily, a problem that is exacerbated when the degree array is very large.
\looseness=-1

To alleviate these consequences, we propose three techniques to reduce the memory consumption of and access to the degree arrays.
The first technique is constructing the degree arrays in reference to an induced subgraph instead of the full graph after running reduction rules exhaustively at the root node (Section~\ref{sec:memory-induced}).
The second technique is maintaining bounds on the non-zero entries of the degree array to avoid iterating over vertices with a degree of zero (Section~\ref{sec:memory-bounds}).
The third technique is using small integer datatypes in the degree array when the maximum degree is small (Section~\ref{sec:memory-short}).
We describe these techniques in the rest of this section.

\subsection{Reducing the Graph and Inducing a Subgraph}\label{sec:memory-induced}

In the state-of-the-art GPU solution for vertex cover problems~\cite{yamout2022parallel}, the degree array is constructed in reference to the whole graph.
That is, every vertex in the original graph has an entry in the degree array.
However, we observe that the reduction rules at the root node have the potential to remove a large number of vertices, either by including them in the solution or by covering all their edges and isolating them.
These vertices have a degree of zero throughout the entire search tree exploration, which means their degrees are stored and checked unnecessarily.
Taking advantage of this observation, we use the CPU to run the reduction rules exhaustively at the root node of the search tree before branching.
We then induce a subgraph on the remaining vertices, and this induced subgraph becomes the basis for branching using the GPU.
The degree arrays are thus constructed with respect to this induced subgraph, making them much smaller.

To further increase the effectiveness of this optimization, we apply the \textit{crown} reduction rule~\cite{chlebik2008crown} at the root node on the CPU as well.
This rule is a sophisticated and heavyweight rule to apply, and our experiments showed us that applying it at every node instead of just the root node may incur more overhead than benefit.
However, applying it just at the root node contributes to further reducing the graph before inducing a subgraph, which assists in further shrinking the size of the degree array.

Applying the reduction rules on the CPU also has another benefit for reducing the memory footprint besides shrinking the size of the degree array.
Removing more vertices at the root allows us to put a tighter bound on the maximum depth of the search tree, and by extension, the maximum depth of the stack that needs to be allocated for each thread block.
Hence, by reducing both the size of the degree array (i.e., the stack entry) and the depth of the stack, we reduce the overall size of the stack that is needed by each thread block.
This optimization enables launching more thread blocks to explore the search tree concurrently for large graphs.

\subsection{Placing Bounds on Non-zero Entries}\label{sec:memory-bounds}

In the state-of-the-art GPU solution for vertex cover problems~\cite{yamout2022parallel}, all vertices in the degree array are checked whenever a reduction rule is applied.
However, deep in the search tree, many of these vertices have a degree of zero and are checked unnecessarily.
An alternative approach is to use a sparse representation of the degree array such as a compacted list of non-zero-degree vertices.
However, such a list would consume additional memory and would be expensive to maintain.

Alternatively, we propose to maintain a bound on the non-zero entries in the degree array and to process vertices only within that bound when applying the reduction rules.
This approach has little memory overhead because it only requires storing two integer indices for the first and last vertex with non-zero degree.
It is also much cheaper to compute the updated bounds at each tree node than to perform a compaction operation.
The trade-off is that we will still check zero-degree vertices within the bounds.
While this trade-off may be expensive in pathological cases, it is an acceptable trade-off in practice and our evaluation shows that the technique is quite effective.

\subsection{Using Small Integer Datatypes}\label{sec:memory-short}

In the state-of-the-art GPU solution for vertex cover problems~\cite{yamout2022parallel}, the degree array is allocated as an array of 32-bit integers.
However, the maximum value that any of these integers may have is the maximum degree in the graph.
If this maximum degree is small enough, we instead use smaller integer datatypes to represent them, thereby further decreasing the size of the degree array in memory.
We exercise care when performing atomic operations on these small integers since atomic operations on small integer datatypes are not natively supported on GPUs.

This optimization is particularly assisted by applying the reduction rules at the root node as mentioned in Section~\ref{sec:memory-induced}.
The reduction rules, particularly the high-degree rule and crown rule, tend to remove vertices with very high degrees.
Hence, applying these rules can substantially reduce the maximum degree in the graph, increasing the chances of using shorter integer datatypes to represent the degree array.

\section{Evaluation}\label{sec:eval}

\begin{table*}
    \centering
    \caption{Execution time (in seconds) of our MVC solution for different graph datasets compared to different baselines}\label{tab:performance}
    \resizebox{\textwidth}{!}{
        
\begin{tabular}{|c|c|c|c|c|c|c|c|c|c|}
    \hline
    \multirow{2}{*}{\textbf{Graph}} & \multirow{2}{*}{\textbf{$|V|$}} & \multirow{2}{*}{\textbf{$|E|$}} & \multicolumn{1}{c|}{\textbf{Yamout}} & \multicolumn{3}{c|}{\textbf{Optimized (component-aware)}} & \multicolumn{3}{c|}{\textbf{Speedup of proposed implementation over}}\\ \cline{5-10}
    & & & \textbf{et al.~\cite{yamout2022parallel}} & \textbf{Sequential} & \textbf{No load balance} & \textbf{Load balanced} {\footnotesize (proposed)} & \textbf{Yamout et al~\cite{yamout2022parallel}} & \textbf{Sequential} & \textbf{No load balance}\\ \hline
    web-webbase-2001~\cite{nr} & 16,062 & 25,593 & $>$6hrs & 0.348 & 0.286  & \textbf{0.007}&  $>$3,085,714$\times$& 49.7$\times$ & 40.9$\times$ \\ \hline    
    power-eris1176~\cite{nr} & 1,176 & 8,688 & $>$6hrs & 2.586 & 0.780 & \textbf{0.040}& $>$540,000$\times$ & 64.7 $\times$& 19.5$\times$\\ \hline    
    movielens-100k\_rating~\cite{nr} & 2,625 & 94,834 & 0.131 & 2.832 & 24.466 & \textbf{0.066}&  1.98$\times$ & 42.9$\times$ & 370.7$\times$ \\ \hline
    qc324~\cite{nr} & 324 & 13,203 & 0.203 & 2.424 & 15.760 & \textbf{0.074}&  2.7$\times$ & 32.7$\times$ & 212.9$\times$ \\ \hline
    SYNTHETIC~\cite{nr} & 30,000 & 58,800 & $>$6hrs & 0.216 & 1.189 & \textbf{0.107} & $>$201,869$\times$ & 2.01$\times$ & 11.1$\times$\\ \hline

    SYNTHETICnew~\cite{nr} & 30,000 & 58,875 & $>$6hrs & 0.230 & 1.170 & \textbf{0.123} & $>$175,609 $\times$ & 1.9$\times$ & 9.5$\times$ \\ \hline
    vc-exact-017~\cite{pace2019} & 23,541 & 34,233 & $>$6hrs & 0.684 & 1.262 & \textbf{0.127} &  $>$170,078$\times$ & 5.4$\times$ & 9.9$\times$ \\ \hline    
    vc-exact-029~\cite{pace2019} & 13,431 & 16,234 & $>$6hrs & 1.730 & 24.905& \textbf{0.133} & $>$1,624,060$\times$ & 13$\times$ & 187.3$\times$ \\ \hline
    c-fat500-5~\cite{nr} & 500 & 23,191 & $>$6hrs & 118.829 & 2,228.442 & \textbf{2.147}&  $>$10,060.5$\times$& 55.3$\times$ & 1,037.9$\times$ \\ \hline   
    scc-infect-dublin~\cite{nr} & 10,972 & 175,573 & $>$6hrs & 780.970 & 633.175 & \textbf{8.937} & $>$2,417$\times$ & 87.4$\times$ & 70.8$\times$ \\ \hline
    rajat28~\cite{nr} & 87,190 & 263,606 & $>$6hrs & 2,799.7& 6,071.161 & \textbf{29.475} & $>$732.8$\times$ & 95$\times$ & 206$\times$ \\ \hline
    rajat20~\cite{nr} & 86,916 & 262,648 & $>$6hrs & 2,768.461 & 6,161.485 & \textbf{30.071} & $>$718.3 $\times$& 92 $\times$ &  204.9$\times$ \\ \hline
    mhda416~\cite{nr} & 416 & 5,177 & 70.5 & 391.111 & 3,705.704 & \textbf{30.6}&  2.3$\times$ & 12.8$\times$ & 121.1$\times$ \\ \hline
    rajat17~\cite{nr} & 94,294 & 277,444 & $>$6hrs & 2,780.674 & 9,696.627 & \textbf{48.273} & $>$447.5 $\times$ & 57.6 $\times$ & 200.87$\times$ \\ \hline
    rajat18~\cite{nr} & 94,294 & 270,253 & $>$6hrs & 2,955.345 & 10,767.830 & \textbf{53.4} & $>$404.5 $\times$ & 55.3 $\times$ & 201.6$\times$\\ \hline    
    web-spam~\cite{nr} & 4,767 & 37,375 & $>$6hrs & 5.628hrs & $>$6hrs & \textbf{152.514} & $>$141.6$\times$ & 132.8$\times$ &  $>$141$\times$ \\ \hline    
    PROTEINS-full~\cite{nr} & 43,471 & 81,044 & $>$6hrs & 5,881.137 & 11,271.386 & \textbf{1,340.876} & $>$16.1$\times$ & 4.386 $\times$ & 8.4$\times$ \\ \hline
\end{tabular}

    }
\end{table*}

\subsection{Experimental Setup}

We implement our code using C++ and CUDA, and compile it with \texttt{nvcc} from the CUDA SDK version 11.7.
We evaluate our CPU implementation using an AMD EPYC 7551P CPU with 128GB of main memory.
We evaluate our GPU implementations using a Volta V100 GPU with 32GB of device memory.
The graph datasets we use in our evaluation are obtained from the Network Data Repository~\cite{nr} and the PACE 2019 Parameterized Algorithms and Computational Experiments Challenge~\cite{pace2019}.
These datasets are listed in Table~\ref{tab:performance}.
We select datasets that take a non-trivial amount of time to run and that are not completely solved by running the reduction rules at the root node of the search tree without branching.
We remove self-loops from the datasets to ensure that they are simple graphs.

\subsection{Performance Comparison}

Table~\ref{tab:performance} compares the execution time of our proposed MVC solution to that of three different baselines for a number of graph datasets that are difficult to solve.

\subsubsection{Comparison with state-of-the-art GPU solution}\label{sec:eval-soa}

The first baseline we compare to in Table~\ref{tab:performance}, Yamout et al.~\cite{yamout2022parallel}, is the state-of-the-art GPU solution for minimum vertex cover.
Our evaluation shows that for most of the graphs we evaluate on, this state-of-the-art solution takes more than 6~hours to finish.
In contrast, our proposed solution significantly outperforms the state-of-the-art solution, yielding speedups with multiple orders of magnitude.
These speedups are attributed to our proposed optimizations of branching on components independently, reducing the graph and inducing a subgraph, and bounding the non-zero entries of the degree arrays.
We evaluate the individual impact of each of these optimizations to the overall speedup in Section~\ref{sec:eval-opt}.

\subsubsection{Comparison with sequential baseline}\label{sec:eval-seq}

The second baseline we compare to in Table~\ref{tab:performance}, \textit{sequential}, is a sequential CPU implementation used as a baseline by prior work~\cite{yamout2022parallel}, which we have enhanced to embody all the optimizations applied to our proposed GPU solution, including component-awareness, shrinking of degree arrays, and bounding of non-zero entries.
The objective of comparing to this baseline is to isolate the improvements due to optimizations from the improvements due to GPU acceleration.
Our results show that our proposed GPU solution outperforms the sequential baseline by one to two orders of magnitude in most cases.
This result shows that even after applying the optimizations, GPUs can still provide promising performance improvements for such branch-and-bound algorithms.
For the cases where the speedup is modest, we investigate the cause in Section~\ref{sec:eval-opt}.

\subsubsection{Importance of load balancing}\label{sec:eval-lb}

The third baseline we compare to in Table~\ref{tab:performance}, \textit{no load balance}, is an intermediate GPU implementation that we provide that embodies all our proposed optimizations, but does not use a worklist and component branch registry for load balancing.
In other words, this implementation assigns thread blocks to different sub-trees and has each thread block explore its sub-tree independently.
If the thread block encounters components, it pushes all the components on its private stack and solves the components independently one after the other, then performs the post-processing itself.
The results show that our proposed solution significantly outperforms this intermediate implementation, highlighting the importance of retaining load balancing capabilities when adding support for branching on components.

\subsection{Impact of Optimizations}\label{sec:eval-opt}

To isolate the benefit of each of our proposed optimizations, Table~\ref{tab:optimizations} evaluates the performance of our proposed GPU solution with and without each of the optimizations enabled.

\begin{table}
    \centering
    \caption{Incremental impact of each optimization}\label{tab:optimizations}
    \resizebox{\columnwidth}{!}{
        
\begin{tabular}{|c|c|c|c|c|}
    \hline
    ~ & \multicolumn{3}{c|}{\textbf{Disabled optimization}} & ~ \\ \cline{2-4}
    \textbf{Graph} & \textbf{Branching on} & \textbf{Reducing and} & \textbf{Bounding non-}&\textbf{Proposed} \\ 
    ~ & \textbf{components} & \textbf{inducing subgraph} & \textbf{zero entires} & ~ \\ \hline

    web-webbase-2001 &  $>$6hrs & 0.04 & 0.009 &\textbf{0.007}\\ \hline
    power-eris1176   &  $>$6hrs & 0.065 & 0.050 & \textbf{0.040}\\ \hline
    movielens-100k\_rating & \textbf{0.059}  & 0.173 & 0.067 &0.066\\ \hline
    qc324 &  0.423 & \textbf{0.029} & 0.043&0.074\\ \hline
    SYNTHETIC   &   $>$6hrs  &0.200 & 0.146 & \textbf{0.107}\\ \hline
    SYNTHETICnew &  $>$6hrs &0.207 & 0.162 & \textbf{0.123}\\ \hline
    vc-exact-017 &  $>$6hrs & 0.227 & 0.136 & \textbf{0.127}\\ \hline
    vc-exact-029 &   $>$6hrs &0.183 & 1.034 & \textbf{0.133}\\ \hline
    c-fat500-5 &  $>$6hrs & 1.354 & \textbf{0.914}&2.147\\ \hline    
    scc-infect-dublin   &  $>$6hrs & 34.156 & 10.345 & \textbf{8.937}\\ \hline
    rajat28       &   $>$6hrs & $>$6hrs &  63.144 &\textbf{29.475}\\ \hline
    rajat20       &  $>$6hrs & $>$6hrs  & 101.721 & \textbf{30.071}\\ \hline
    mhda416 & 87.694  & \textbf{30.4} & 38.843 &30.6\\ \hline
    rajat17       &  $>$6hrs & $>$6hrs& 164.148 & \textbf{48.273}\\ \hline    
    rajat18      &   $>$6hrs & $>$6hrs & 137 & \textbf{53.4}\\ \hline
    web-spam      &  $>$6hrs & 3,522.894 &  990& \textbf{152.514} \\ \hline    
    PROTEINS-full &  $>$6hrs & $>$6hrs & 7,332.105 & \textbf{1,340.876}\\ \hline
\end{tabular}

    }
\end{table}

\subsubsection{Branching on components}\label{sec:eval-components}

We observe from Table~\ref{tab:optimizations} that when we disable branching on components, most of the graphs take longer than 6~hours to solve.
This observation shows that branching on components is an essential optimization on its own that contributes substantially to the overall performance improvement.
To better understand the benefit of branching on components, Table~\ref{tab:component} shows the number of search tree nodes visited when branching on components is disabled compared to the number of search tree nodes visited by our proposed GPU solution.
It also shows how many of the search tree nodes branch on components and a histogram of the number of components they branch on.
For runs that exceed 6~hours, the number reported is the number of search tree nodes visited before the run is killed.
It is clear that with branching on components enabled, the search tree nodes branch on components frequently sometimes yielding many components, which substantially reduces the number of search tree nodes visited and explains the reduction in execution time.
The only graph in Table~\ref{tab:optimizations} where branching on components does not benefit performance is \textit{movielens-100k\_rating}.
This graph is also the only graph in Table~\ref{tab:component} that does not experience a reduction in the number of tree nodes visited because it does not branch on components frequently, and when it does, the number of components is always small.

\begin{table*}
    \centering
    \caption{Total search tree nodes visited by our proposed GPU solution without and with branching on components enabled}\label{tab:component}
    \resizebox{\textwidth}{!}{

\begin{tabular}{|c|c|c|c|c|}
    \hline
    \multirow{2}{*}{\textbf{Graph}} & \textbf{Branching on components disabled} & \multicolumn{3}{c|}{\textbf{Branching on components enabled (Proposed)}}\\\cline{2-5}
    & \textbf{Tree nodes visited} & \textbf{Tree nodes visited} & \textbf{Branches on components} & \textbf{Components per branch} \{ number of components: frequency \} \\ \hline
    web-webbase-2001 & $>$393,583,034,052 & 25,668 & 1,686 & \{2: 1,272; 3: 311; 4: 54; 5: 16; 6: 4; 12: 16; 13: 6; 14: 1; 15: 2; 18: 2; 21: 2\} \\ \hline
    power-eris1176 & $>$1,332,311,399 & 958,769 & 162,086 & \{2: 158,176; 3: 3,909; 4: 1\} \\ \hline
    movielens-100k\_rating & 140,149  & 140,731 & 339 & \{2: 337; 3: 2\} \\ \hline
    qc324 & 3,721,547 & 183,486 & 5,426 & \{2: 5,426\} \\ \hline
    SYNTHETIC & $>$3,744,064,626 & 22,205 & 1 & \{300: 1 \} \\ \hline
    SYNTHETICnew & $>$4,988,093,891 & 21,482 & 180 & \{2:  178; 3:  1; 300:  1\} \\ \hline
    vc-exact-017 & $>$61,020,066,119 & 15,860 & 2,305 & \{2: 1,915; 3: 345; 4: 41; 5: 2; 6: 1; 110: 1\} \\ \hline
    vc-exact-029 & $>$51,361,041,488 & 455,942 & 55,895 & \{2: 49,933; 3: 5,386; 4: 515; 5: 60; 91: 1\} \\ \hline
    c-fat500-5 & $>$446,159,772,984  &  11,989,943 & 2,823,500 & \{2: 2,823,500 \} \\ \hline
    scc-infect-dublin & $>$30,556,645,527 & 16,947,293 & 1,782,586 & \{2: 1,737,241; 3: 45,292; 4: 52; 119: 1\} \\ \hline
    rajat28 & $>$105,502,762,711 &384,874,435 & 38,810,400 & \{2: 38,461,399; 3: 344,207; 4: 1,654; 5: 566; 6: 2,569; 7: 4; 165: 1\} \\ \hline
    rajat20 & $>$152,025,008,677 & 386,531,313 & 38,868,775 & \{2: 38,519,806; 3: 344,291; 4: 1,641; 5: 468; 6: 2,568; 172: 1\} \\ \hline
    mhda416 & 1,979,281,153  &  690,193,632 & 59,684,057 & \{2: 59,679,037; 3: 5,020\} \\ \hline
    rajat17 & $>$108,721,552,326 & 627,349,654 & 64,238,954 & \{2: 63,742,743; 3: 489,672; 4: 3,310; 5: 556; 6: 2,672; 178: 1\} \\ \hline
    rajat18 & $>$80,830,011,366 & 633,894,898 & 65,650,458 & \{2: 64,735,163; 3: 646,532; 4: 144,638; 5: 105,705; 6: 18,419; 165: 1\} \\ \hline    
    web-spam & $>$64,976,746,153 & 1,561,594,405 & 221,403,914 & \{2: 165,762,811; 3: 44,676,907; 4: 9,421,137; 5: 1,316,802; 6: 190,218; 7: 30,881; 8: 4,379; 9: 645; 10: 127; 11: 7\} \\ \hline    
    PROTEINS-full & $>$1,144,308,917 & 166,043,465 & 16,973,316 & \{2: 15,033,238; 3: 1,784,772; 4: 152,095; 5: 3,199; 6: 11; 1,168: 1\} \\ \hline
\end{tabular}

    }
\end{table*}

\subsubsection{Reducing the graph and inducing a subgraph}\label{sec:eval-memory}

We observe from Table~\ref{tab:optimizations} that when we disable the optimization of reducing the graph and inducing a subgraph at the root node of the search tree, it can increase latency significantly, with a number of graphs taking longer than 6~hours to solve.
This observation shows that reducing the graph and inducing a subgraph is also an essential optimization on its own that is essential for the performance improvement achieved by our proposed GPU solution.
To better understand the benefit of reducing the graph and inducing a subgraph, Table~\ref{tab:memory} shows the number of vertices in the degree array before and after this optimization, and the impact on the number of thread blocks executed.
The table also shows whether or not the degree array fits in shared memory.
It is clear that constructing the degree array from the induced subgraph after reduction substantially reduces the number of vertices it contains, and that our memory optimizations enable more thread blocks to execute in parallel, and more graphs to have their degree arrays fit in shared memory.
The only graph in Table~\ref{tab:optimizations} where reducing the graph and inducing a subgraph impacts performance negatively is \textit{qc324}.
This graph is also the only graph in Table~\ref{tab:memory} that does not experience an increase in the number of thread blocks launched because it was already at the maximum number.

\begin{table*}
    \centering
    \caption{Impact of reducing the graph and inducing a subgraph on the degree array, blocks launched, and use of shared memory}\label{tab:memory}
    \resizebox{0.7\textwidth}{!}{
        
\begin{tabular}{|c|c|c|c|c|c|c|c|c|c|c|}
    \hline
    \multirow{3}{*}{\textbf{Graph}} & \multicolumn{3}{c|}{\textbf{Number of vertices in the}} & \multicolumn{3}{c|}{\textbf{Number of thread blocks}} & \multicolumn{2}{c|}{\textbf{Degree array fits}}& \multicolumn{2}{c|}{\textbf{Short data type for}}  \\
     & \multicolumn{3}{c|}{\textbf{degree array}} & \multicolumn{3}{c|}{\textbf{launched}} & \multicolumn{2}{c|}{\textbf{in shared memory}} & \multicolumn{2}{c|}{\textbf{degree array}} \\ \cline{2-11}
    & \textbf{Before} & \textbf{After} & \textbf{Ratio} & \textbf{Before} & \textbf{After} & \textbf{Increase} & \textbf{Before} & \textbf{After}& \textbf{Before} & \textbf{After}\\ \hline
    web-webbase-2001  & 16,062 & 1,631 & 0.10$\times$ & 71 & 640 & 9.01$\times$ & No & Yes & No & Yes \\ \hline
    power-eris1176  & 1,176 & 428 & 0.36$\times$ & 640 & 2,560 & 4.00$\times$ &Yes & Yes & No & Yes\\ \hline
    movielens-100k\_rating  & 2,625 & 768 & 0.29$\times$ & 320 & 1,280 & 4$\times$ & Yes & Yes& No & Yes \\ \hline 
    qc324  & 324 & 324 & 1$\times$ & 2,560 & 2,560 & 1$\times$ & Yes & Yes & No & Yes\\ \hline
    SYNTHETIC       & 30,000 & 21,900 & 0.73$\times$ & 6 & 39 & 6.5$\times$ & No & No & No & Yes \\ \hline
    SYNTHETICnew & 30,000 & 22,208 & 0.74$\times$ & 6 & 37 & 6.17$\times$ & No & No & No & Yes\\ \hline
    vc-exact-017    & 23,541 & 5,718 & 0.24$\times$ & 3 & 320 & 106.7$\times$ & No & Yes& No & Yes\\ \hline
    vc-exact-029 & 13,431 & 5,858 & 0.43$\times$ & 2 & 320 & 160$\times$ & No & Yes & No & Yes\\ \hline
    c-fat500-5 & 500 & 500 & 1$\times$ & 1,280 & 2,560 & 2$\times$ & Yes & Yes& No & Yes \\ \hline
    scc-infect-dublin & 10,972 & 9,785 & 0.89$\times$ & 14 & 124 & 8.86$\times$ & No & Yes & No & Yes\\ \hline
    rajat28     & 87,190 & 3,455 & 0.04$\times$ & 1 & 320 & 320$\times$ & No & Yes & No & Yes\\ \hline
    rajat20     & 86,916 & 3,628 & 0.04$\times$ & 1 & 320 & 320$\times$ & No & Yes & No & Yes\\ \hline
    mhda416  & 416 & 409 & 0.98$\times$ & 1,280 & 2,560 & 2$\times$ & Yes & Yes & No & Yes\\ \hline
    rajat18     & 94,294 & 4,549 & 0.05$\times$ & 1 & 320 & 320$\times$ & No & Yes & No & Yes \\ \hline
    rajat17     & 94,294 & 3,476 & 0.04$\times$ & 1 & 320 & 320$\times$ & No & Yes& No & Yes \\ \hline
    
    web-spam    & 4,767 & 843 & 0.18$\times$ & 160 & 1,280 & 8.00$\times$ & Yes & Yes & No & Yes\\ \hline
    PROTEINS-full & 43,471 & 36,099 & 0.83$\times$ & 1 & 11 & 11.00$\times$ & No & No & No & Yes\\ \hline
\end{tabular}

    }
\end{table*}

\subsubsection{Placing bounds on non-zero entries}\label{sec:eval-bound}

We observe from Table~\ref{tab:optimizations} that when we disable the optimization of placing bounds on the non-zero entries of the degree arrays, it can modestly increase latency.
Although the impact of this optimization is not as major as the impact of the other optimizations, it still provides a substantial benefit across the majority of graphs.

\subsection{Performance Breakdown}\label{sec:eval-breakdown}

\begin{figure}
    \centering
    \includegraphics[width=\columnwidth]{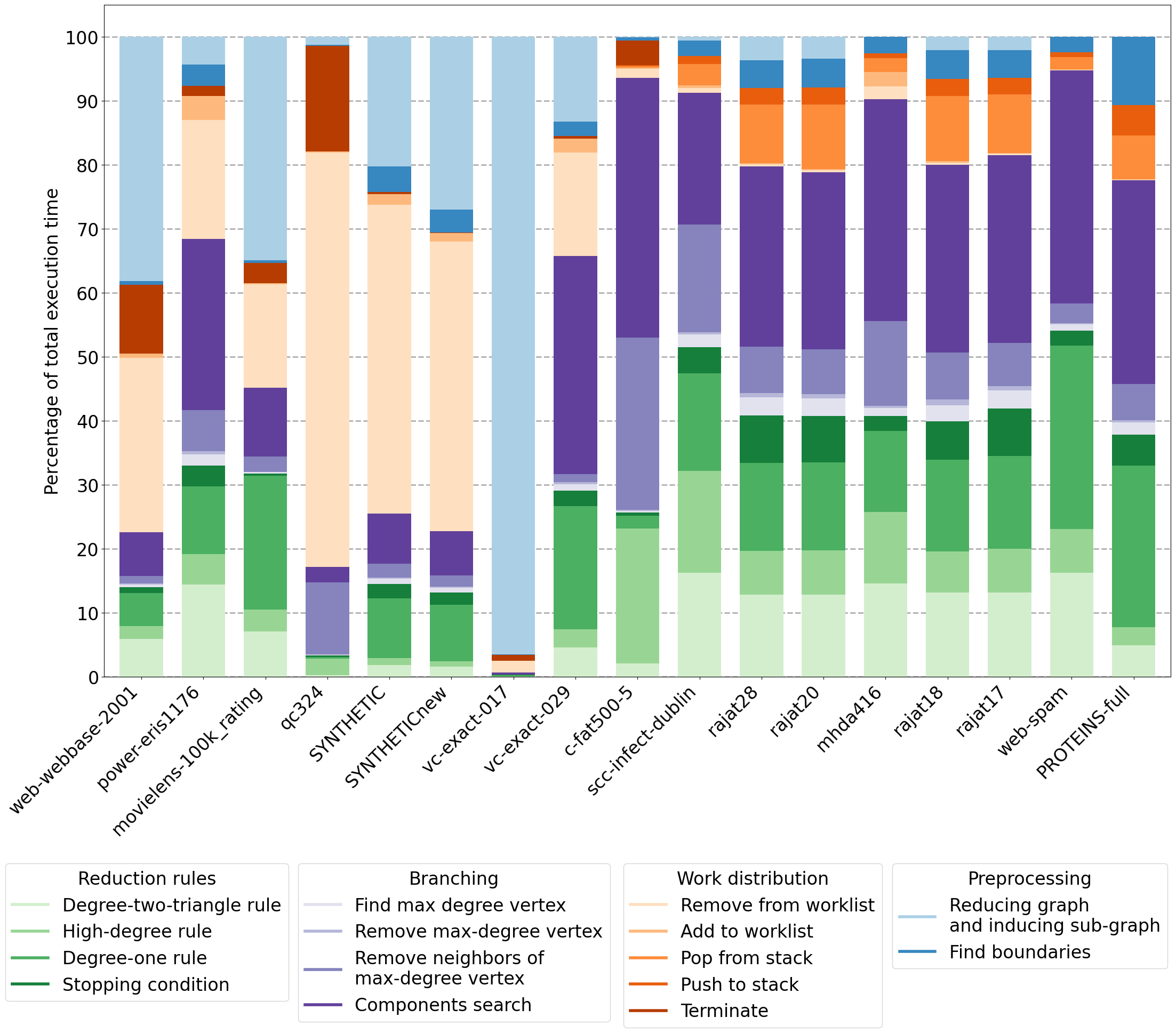}
    \caption{Breakdown of execution time for our proposed MVC solution}\label{fig:breakdown}
\end{figure}

Figure~\ref{fig:breakdown} shows how the execution time of our proposed implementation is broken down across different activities.
The CPU pre-processing time for applying reduction rules and inducing a subgraph is included in the \textit{reducing graph and inducing subgraph} activity.
The breakdown of the GPU time for the remaining activities is obtained by instrumenting the code using SM clocks to count how many cycles each thread block spends in each activity, then normalizing the counts per thread block and taking their mean across thread blocks.
The breadth first search operation to find components and the atomic operations to update the registry are included in the \textit{components search} activity.

It is clear from Figure~\ref{fig:breakdown} that for the graph datasets with long execution time, little time is spent accessing the stack and worklist data structures, which demonstrates the efficiency of our load balancing scheme.
A substantial component of the execution time is spent applying the reduction rules, which is the most productive part of the computation.
Additionally, finding the components takes a notable percentage of the execution time, however, this penalty is worth paying because of the major improvement we get from branching on components independently.

\subsection{Parameterized Vertex Cover}\label{sec:eval-pvc}

Table~\ref{tab:pvc} compares the execution time of our proposed PVC solution to the same baselines and for the same graph datasets used for evaluating MVC.
For each graph dataset, we evaluate three instances where the parameter $k$ is equal to the minimum, minimum - 1, and minimum + 1.
Our overall observations are the same as those for MVC: our proposed GPU solution mostly outperforms the state-of-the-art GPU solution, and sequential CPU solution with the same optimizations that we propose, and a GPU solution with the same optimizations that we propose but that does not apply load balancing.

\begin{table*}
    \centering
    \caption{Execution time (in seconds) of our PVC solution for different graph datasets and values of $k$ compared to different baselines}\label{tab:pvc}
    \resizebox{\textwidth}{!}{
        
\begin{tabular}{|c|c|c|c|c|c|c|c|c|}
    \hline
    \multirow{2}{*}{\textbf{Graph}} & \multirow{2}{*}{\textbf{Instance}} & \multicolumn{1}{c|}{\textbf{Yamout}} & \multicolumn{3}{c|}{\textbf{Optimized (component-aware)}} & \multicolumn{3}{c|}{\textbf{Speedup of proposed implementation over}}\\ \cline{4-9}
    & & \textbf{et al.~\cite{yamout2022parallel}} & \textbf{Sequential} & \textbf{No load balance} & \textbf{Load balanced} {\footnotesize (proposed)} & \textbf{Yamout et al~\cite{yamout2022parallel}} & \textbf{Sequential} & \textbf{No load balance}\\ \hline
    
    \multirow{3}{*}{web-webbase-2001} & \small{k = min-1} & $>$2hrs & 0.316 & 0.216 & \textbf{0.008} & $>$900,000$\times$ & 39.5$\times$ & 27$\times$\\
    \cline{2-9}
                             & \small{k = min}  & $>$2hrs & 0.316 & 0.009 & \textbf{0.006} & $>$1,200,000$\times$& 52.7$\times$ & 1.5$\times$\\
    \cline{2-9}
                             & \small{k = min+1} & 0.060 & 0.315 &0.009 & \textbf{0.006} & 10$\times$ & 52.5$\times$& 1.5$\times$\\
    \hline
    \multirow{3}{*}{power-eris1176} & \small{k = min-1} & $>$2hrs & 2.246 & 0.699 & \textbf{0.041} & $>$175,610$\times$ & 55$\times$ & 17$\times$\\
    \cline{2-9}
                             & \small{k = min} & \textbf{0.008} & 2.219 & 0.028& 0.035 & 0.23$\times$ & 63.4$\times$& 0.8$\times$ \\
    \cline{2-9}
                             & \small{k = min+1} & \textbf{0.005} & 2.216 & 0.019  & 0.015 & 0.33$\times$ & 147.7$\times$& 1.3$\times$ \\
    \hline
    \multirow{3}{*}{movielens-100k\_rating} & \small{k = min-1} & 0.130 & 2.802 & 18.383 & \textbf{0.068} & 1.9$\times$ & 41.2$\times$ & 270.3$\times$\\
    \cline{2-9}
                             & \small{k = min}  & \textbf{0.021} & 0.036& 0.037 & 0.055 & 0.38$\times$& 0.65$\times$ & 0.67$\times$\\
    \cline{2-9}
                             & \small{k = min+1} & \textbf{0.021} & 0.036 & 0.042 & 0.060 & 0.35$\times$& 0.6$\times$& 0.7$\times$\\
    \hline
    
    \multirow{3}{*}{qc324} & \small{k = min-1} & 0.194 & 2.464 & 14.723 & \textbf{0.082} & 2.36$\times$ & 30$\times$ & 179.5$\times$\\
    \cline{2-9}
                             & \small{k = min}  & \textbf{0.002} & 2.412 & 0.005 & 0.005 & 0.4$\times$& 482.4$\times$ & 1$\times$\\
    \cline{2-9}
                             & \small{k = min+1} & \textbf{0.001} & 2.495 & 0.005 & 0.004 & 0.25$\times$& 623.8$\times$& 1.25$\times$\\
    \hline
    \multirow{3}{*}{SYNTHETIC} & \small{k = min-1} & $>$2hrs & 0.184 & 1.201 & \textbf{0.109} & $>$66,055$\times$& 1.7$\times$& 11$\times$\\
    \cline{2-9}
                             & \small{k = min}  & 0.540 & 0.183 & 1.142 & \textbf{0.108} & 5$\times$& 1.7$\times$ & 10.6$\times$\\
    \cline{2-9}
                             & \small{k = min+1} & 0.560 & 0.185 & 1.137 & \textbf{0.108} & 5.2$\times$ & 1.7$\times$& 10.5$\times$\\
    \hline
    \multirow{3}{*}{SYNTHETICnew} & \small{k = min-1} & $>$2hrs & 0.266 & 1.184 & \textbf{0.122} & $>$59,016$\times$& 2.2$\times$ & 9.7$\times$\\
    \cline{2-9}
                             & \small{k = min} & $>$2hrs & 0.227 & 1.123 & \textbf{0.122} & $>$59,016$\times$& 1.9$\times$& 9.2$\times$\\
    \cline{2-9}
                             & \small{k = min+1} & $>$2hrs & 0.253 & 1.124 & \textbf{0.122} & $>$59,016$\times$& 2$\times$& 9.2$\times$\\
    \hline
    \multirow{3}{*}{vc-exact-017} & \small{k = min-1} & $>$2hrs & 0.686 & 1.088 & \textbf{0.138} & $>$52,174$\times$& 5$\times$& 7.9$\times$\\
    \cline{2-9}
                             & \small{k = min} & $>$2hrs & 0.692 & 1.037 & \textbf{0.138} & $>$52,174$\times$& 5$\times$& 7.5$\times$\\
    \cline{2-9}
                             & \small{k = min+1} & $>$2hrs & 0.683 & 0.630 & \textbf{0.137} & $>$52,554$\times$& 5$\times$ & 4.6$\times$\\
    \hline
    \multirow{3}{*}{vc-exact-029} & \small{k = min-1} & $>$2hrs & 1.722 & 23.716 & \textbf{0.131} & $>$54,962$\times$& 13.1$\times$& 181$\times$\\
    \cline{2-9}
                             & \small{k = min} & $>$2hrs & 1.772 & 9.408 & \textbf{0.094} & $>$76,595$\times$& 18.9$\times$ & 100$\times$\\
    \cline{2-9}
                             & \small{k = min+1} & $>$2hrs & 1.843 & 10.785 & \textbf{0.091} & $>$79,121$\times$& 20.3$\times$ & 118.5$\times$\\
    \hline
    \multirow{3}{*}{c-fat500-5} & \small{k = min-1} & $>$2hrs & 121.772 & 1,864.019 & \textbf{2.198} & $>$3,276$\times$ & 55.4$\times$ & 848$\times$\\
    \cline{2-9}
                             & \small{k = min}  & \textbf{0.005} & 122.06 & 0.006 & 0.008 & 0.625$\times$ & 15,258$\times$ & 0.75$\times$\\
    \cline{2-9}
                             & \small{k = min+1} & \textbf{0.002} & 120.221 & 0.007 & 0.007 & 0.3$\times$& 17,174$\times$& 1$\times$\\
    \hline
    \multirow{3}{*}{scc-infect-dublin} & \small{k = min-1} & $>$2hrs & 757.502 & 855.349 &  \textbf{10.803} & $>$666.5$\times$ & 70$\times$& 79$\times$\\
    \cline{2-9}
                             & \small{k = min}   & $>$2hrs & 758.6  & 822.406 & \textbf{6.082} & $>$1,184$\times$& 125$\times$& 135$\times$\\
    \cline{2-9}
                             & \small{k = min+1} & $>$2hrs & 758.805 & 904.985 & \textbf{6.138} & $>$1,173$\times$& 123.6$\times$& 147.4$\times$\\
    \hline
    \multirow{3}{*}{rajat28} & \small{k = min-1} & $>$2hrs & 2,728.663 & $>$2hrs & \textbf{33.906} & $>$212.4$\times$ & 80.5$\times$ & $>$212$\times$\\
    \cline{2-9}
                             & \small{k = min} & $>$2hrs & 2,727.505 & $>$2hrs &  \textbf{16.939} & $>$425$\times$ & 161$\times$ & $>$425$\times$\\
    \cline{2-9}
                             & \small{k = min+1}  & $>$2hrs & 2,719.056 & $>$2hrs &  \textbf{1.158} & $>$6,218$\times$ & 2,348$\times$ & $>$6,218$\times$\\
    \hline
    \multirow{3}{*}{rajat20} & \small{k = min-1} & $>$2hrs & 2,773.723 & $>$2hrs & \textbf{34.527} & $>$208.5$\times$& 80.3$\times$& $>$208.5$\times$\\
    \cline{2-9}
                              & \small{k = min} & $>$2hrs & 2,721.855 & $>$2hrs & \textbf{2.397} & $>$3,004$\times$& 1,135$\times$& $>$3,003$\times$\\
    \cline{2-9}
                              & \small{k = min+1} & $>$2hrs & 2,767.719 & $>$2hrs & \textbf{1.103} & $>$6,528$\times$& 2,509$\times$& $>$6,527$\times$\\
    \hline
    \multirow{3}{*}{mhda416} & \small{k = min-1} & 71.895 & 385.258 & 9,533.311 & \textbf{24.570} & 2.9$\times$ & 15.7$\times$ & 388$\times$\\
    \cline{2-9}
                             & \small{k = min}  & 0.007 & \textbf{0.004} & 0.051 & 0.136 & 0.05$\times$& 0.03$\times$ & 0.375$\times$\\
    \cline{2-9}
                             & \small{k = min+1} & 0.007 & \textbf{0.0004} & 0.057 & 5.200 & 0.001$\times$& 0.00008$\times$& 0.01$\times$\\
    \hline
    \multirow{3}{*}{rajat18} & \small{k = min-1} & $>$2hrs & 2,939.643 & $>$2hrs &  \textbf{61.211} & $>$118$\times$ & 48$\times$&$>$117$\times$\\
    \cline{2-9}
                              & \small{k = min} & $>$2hrs & 3,004.724 & $>$2hrs &  \textbf{1.349} & $>$5,337$\times$ & 2,227$\times$&$>$5,337$\times$\\
    \cline{2-9}
                              & \small{k = min+1} & $>$2hrs & 2,921.240 & $>$2hrs &  \textbf{1.371} & $>$5,252$\times$ & 2,131$\times$& $>$5,251$\times$\\
    \hline
    \multirow{3}{*}{rajat17} & \small{k = min-1} & $>$2hrs & 2,700.941 & $>$2hrs &  \textbf{55.636}& $>$ 129$\times$& 48.5$\times$& $>$129$\times$\\
    \cline{2-9}
                              & \small{k = min} & $>$2hrs & 2,698.708 & $>$2hrs & \textbf{1.536} & $>$4,687.5$\times$& 1,757$\times$& $>$4,687$\times$\\
    \cline{2-9}
                              & \small{k = min+1} & $>$2hrs & 2,737.359 & $>$4hrs &\textbf{1.130}& $>$ 6,372$\times$& 2,422.4$\times$& $>$12,743$\times$\\
    \hline
    
    \multirow{3}{*}{web-spam} & \small{k = min-1} & $>$2hrs & $>$2hrs & $>$2hrs &  \textbf{150.618}  & $>$48$\times$ & $>$48$\times$ & $>$48$\times$\\
    \cline{2-9}
                             & \small{k = min} & \textbf{0.060} & $>$2hrs & 17.429 & 6.101 &  0.009$\times$ & $>$1,180$\times$ & 2.86$\times$ \\
    \cline{2-9}
                             & \small{k = min+1} & \textbf{0.008} & $>$2hrs & 0.017 & 1.587  & 0.005$\times$ & $>$4,537$\times$ & 0.01$\times$ \\
    \hline
    \multirow{3}{*}{PROTEINS-full} & \small{k = min-1} & $>$6hrs & 5,778.315 & $>$2hrs & \textbf{1,298.213} & $>$16.6$\times$ & 4.45$\times$& $>$5.5$\times$\\
    \cline{2-9}
                              & \small{k = min} & $>$6hrs & 5,711.873  & $>$2hrs & \textbf{1,291.068} & $>$16.7$\times$& 4.4$\times$& $>$5.6$\times$\\
    \cline{2-9}
                              & \small{k = min+1} & $>$6hrs & 5,898.726  & $>$2hrs & \textbf{1,243.744} & $>$17.4$\times$& 4.7$\times$& $>$5.8$\times$\\
    \hline
\end{tabular}

    }
\end{table*}

There are only a few of graph datasets where the state-of-the-art GPU solution outperforms our proposed solution.
Recall that the difference between MVC and PVC is that MVC searches exhaustively for the minimum, whereas PVC terminates as soon as it finds a vertex cover that satisfies the parameter.
The cases where the state-of-the-art solution outperforms our proposed solution are the short-running cases where a satisfactory vertex cover is luckily found on an early branch.
However, for the long-running cases where speedup matters more, our proposed solution always outperforms the state-of-the-art solution.

\subsection{Comparison Using Prior Work's Datasets}\label{sec:eval-old-datasets}

Table~\ref{tab:eval-old-datasets} compares the execution time of our proposed solution to that of the state-of-the-art GPU solution~\cite{yamout2022parallel} using the same graph datasets used in that prior work's evaluation.
It is clear that our proposed solution outperforms prior work for most of the non-synthetic graphs.
However, it does not outperform prior work for a set of synthetic graphs from the same family.
These synthetic graphs do not benefit from our optimizations because they are very dense so they do not break into components, and they have a small number of vertices so their memory footprint is already low.

Empirically speaking, density could be used as a hint to predict whether or not our proposed solution would perform better than prior work~\cite{yamout2022parallel}.
Using the combined set of graphs from Table~\ref{tab:performance} and Table~\ref{tab:eval-old-datasets}, 20 out of 21 graphs where our solution is faster have a density of less than 10\%.
Meanwhile, 9 out of 10 graphs where our solution is slower have a density that is greater than 10\%.
However, from a theoretical standpoint, no such guarantee can be made and pathological cases can always be constructed.
To the best of our knowledge, there is no cheap method for predicting whether or not the graph will split into components as it is being solved.

\begin{table}
    \centering
    \caption{Execution time (in seconds) of our MVC solution compared to prior work using the same graph datasets used in that work}\label{tab:eval-old-datasets}
    \resizebox{\columnwidth}{!}{
        \begin{tabular}{|l|c|c|c|}
\hline
\textbf{Graphs} & \textbf{Yamout et al.~\cite{yamout2022parallel}} & \textbf{Proposed} & \textbf{Speed up} \\ \hline
\multicolumn{4}{|l|}{\textbf{Low degree}} \\ \hline
US power grid & 0.852 & \textbf{0.001} & 852$\times$ \\ \hline
Sister Cities & 0.106 & \textbf{0.001} & 106$\times$\\ \hline
LastFM Asia & 0.939 & \textbf{0.005} & 187.8$\times$ \\ \hline
\multicolumn{4}{|l|}{\textbf{High degree}} \\ \hline
movielens-100k rating & 0.132 & \textbf{0.066} & 2$\times$ \\ \hline
wikipedia\_link\_lo & 387.628 & \textbf{0.140} & 2,768$\times$\\ \hline
wikipedia\_link\_csb & \textbf{0.034} & 0.177 & 0.19$\times$ \\ \hline
cop\_hat300-1 & \textbf{0.021} & 0.141 & 0.15$\times$\\ \hline
cop\_hat300-2 & \textbf{0.029} & 0.073 & 0.4$\times$\\ \hline
cop\_hat300-3 & \textbf{1.657} & 2.774 & 0.6$\times$ \\ \hline
cop\_hat500-1 & \textbf{0.092} & 0.497 & 0.2$\times$\\ \hline
cop\_hat500-2 & \textbf{1.558} & 1.763 & 0.88$\times$\\ \hline
cop\_hat500-3 & \textbf{1,018.898} & 1,387.871 & 0.73$\times$\\ \hline
cop\_hat700-1 & \textbf{0.238} & 1.540 & 0.15$\times$\\ \hline
cop\_hat700-2 & \textbf{31.241} & 34.092 & 0.92$\times$\\ \hline
cop\_hat1000-1 & \textbf{1.400} & 2.000 & 0.7$\times$ \\ \hline
cop\_hat1000-2 & 4,527.601 & \textbf{4,513.368} & 1.003 $\times$\\ \hline
\end{tabular}

    }
\end{table}

When in doubt, users can always make the conservative decision of using our proposed solution.
Even when our solution does not outperform prior work, its execution time remains reasonable.
On the other hand, for the datasets used in our main evaluation in Table~\ref{tab:performance} where our solution outperforms prior work, their execution time is unreasonable, exceeding 6~hours most of the time.

\section{Related Work}\label{sec:related}

\textbf{Vertex cover.}
Significant efforts have been made to develop efficient exact algorithms for NP-hard problems like Minimum Vertex Cover (MVC)~\cite{DBLP:books/fm/GareyJ79}. One of the earliest improvements over brute-force search was the $\mathcal{O}(2^{n/3})$-time algorithm by Tarjan and Trojanowski~\cite{DBLP:journals/siamcomp/TarjanT77} for Maximum Independent Set (MIS), which is equivalent to MVC. This was later improved by Robson’s $\mathcal{O}(1.1889^n)$-time algorithm~\cite{DBLP:journals/jal/Robson86, DBLP:journals/jal/RobsonNew} and further refinements~\cite{DBLP:journals/iandc/XiaoN17,DBLP:journals/jco/XiaoN17}. In the Parameterized Vertex Cover (PVC) problem, where the goal is to find a vertex cover of size at most $k$, fixed-parameter tractable (FPT) algorithms exist, beginning with Fellows' $\mathcal{O}(2^k \cdot n)$-time algorithm~\cite{DF97}, later refined through various  techniques~\cite{DBLP:journals/ipl/BalasubramanianFR98,DBLP:conf/wg/ChenKJ99,DBLP:journals/networks/ChenLJ00,DBLP:conf/aaecc/FellowsK93,BussG93}, culminating in Chen et al.’s $\mathcal{O}(1.2738^k + kn)$-time algorithm~\cite{DBLP:journals/tcs/ChenKX10}.
Multiple CPU implementations of vertex cover algorithms exist, including sequential~\cite{pace2019,DBLP:conf/siamcsc/HespeL0S20,DBLP:journals/tcs/AkibaI16} and parallel~\cite{DBLP:journals/algorithmica/Abu-KhzamLSS06,DBLP:journals/jpdc/Abu-KhzamDMN15} ones.
The focus of our work is on parallel GPU-based implementations, which have only recently gained attention.
\looseness=-1

\textbf{Independent set.}
Since the complement of a minimum vertex cover is a maximum independent set, one can compute a maximum independent set and derive a minimum vertex cover.
Exact algorithms for maximum independent set also rely on branching, and graphs may break into components in the process~\cite{alsahafy2019computing,hespe2021targeted}.
Our proposed techniques for load balancing non-tail-recursive parallel branching can also be used in parallel implementations of exact maximum independent set.
Our main contribution is orthogonal to the distinctions between vertex cover and independent set algorithms.

\textbf{GPU solutions for vertex cover problems.}
A number of works use GPUs to parallelize minimum vertex cover or maximum independent set using heuristic non-branching algorithms~\cite{toume2014gpu,burtscher2018high,imanaga2020efficient,mongandampulath2025multi,waldherr2025maximum}.
In our work, we focus on exact algorithms which require parallelizing a branch-and-reduce search tree exploration.
The prior works accelerating the exact branch-and-reduce algorithm~\cite{jufinding,abu2018accelerating,kabbara2013parallel,yamout2022parallel} do so by exploring different sub-trees in parallel.
Liu et al.~\cite{jufinding} explore the top portion of the tree on CPUs and delegate deeper sub-trees to GPUs, but this approach performs frequent communication between the CPU and the GPU and launches many small grids, which is inefficient.
Our approach uses a single-kernel solution.
Single-kernel solutions have been proposed~\cite{abu2018accelerating,kabbara2013parallel,yamout2022parallel} and are discussed in Section~\ref{sec:background}.
Briefly, these solutions assign different thread blocks to different sub-trees, with the state-of-the-art solution~\cite{yamout2022parallel} also focusing on load balance.
However, these solutions struggle with large and complex graphs because they are not component-aware and have a high memory footprint, two limitations that our proposed techniques overcome.
Although our techniques can be ported to parallel vertex cover algorithms for CPUs, there are key distinctions between CPUs and GPUs that may affect this portability.
First, CPUs have fewer threads which makes them less sensitive to load imbalance, thereby not requiring the load balancing technique to be as aggressive.
Second, CPU threads have large stacks so they are more likely to traverse the search tree using function recursion than to use a programmer-managed stack.
Third, CPU threads can easily spawn new threads dynamically when they discover more work to assist with load balancing, whereas dynamic parallelism on GPUs has high performance overhead~\cite{el2016klap,tang2017controlled,olabi2022compiler} which motivates the explicit offloading mechanism that we propose.

\textbf{Branching algorithms on GPUs.}
Beyond vertex cover, branching algorithms for graph problems have been widely studied on GPUs in recent years.
GPU solutions have been proposed for many problems such as pattern mining~\cite{chen2020pangolin,chen2021sandslash,kawtikwar2023beep}, maximal clique enumeration~\cite{jenkins2011lessons,wei2021accelerating,almasri2023parallelizing}, and $k$-clique counting~\cite{almasri2022parallel}.
However, all these prior works parallelize search tree exploration in the context of tail-recursive branches which are naturally amenable to load balancing.
For example, when branching from a clique to the larger cliques that contains it, we can continue exploring each of the larger cliques independently.
There is no post-processing that needs to be done at the level of the smaller clique when the exploration of the larger cliques has completed.
This tail-recursive pattern makes it easy to offload the processing of the larger cliques to other thread blocks.
In contrast, our work tackles the parallel exploration of non-tail-recursive branches where load balancing is not as trivial.

\textbf{Graph algorithms on GPUs.}
GPUs have been widely used to accelerate numerous graph algorithms~\cite{shi2018graph,harish2007accelerating,vineet2008cuda,merrill2012scalable,merrill2015high,wu2013complexity,pan2017multi,yin2022glign}.
There are a number of libraries for performing graph computations on GPUs~\cite{10.1145/3173162.3173180,wang2016gunrock,kang2023cugraph}.
However, these libraries mostly support polynomial time algorithms where the main parallelism comes from iterating over vertices of edges in parallel.
In contrast, our work is concerned with a different class of graph algorithms that follow the branch-and-reduce paradigm.
While our implementation uses block-level routines that iterate over vertices or edges in parallel, such as our breadth-first-search routine for finding components, the dominant pattern in our work is parallelizing the traversal of the search tree, which is not the focus of these libraries.
\looseness=-1

\section{Conclusion}\label{sec:conclusion}

We proposed a novel GPU-based approach for solving the minimum vertex cover and parameterized vertex cover problems.
Our solution branches on graph components independently to reduce redundant computations.
Since the resulting branching pattern is non-tail-recursive which interferes with load balancing, we devise a technique to track active descendants of a branch and delegate the post-processing to the last descendant.
Our solution thus retains the load balancing capabilities and supports multiple nesting of non-tail-recursive branches.
Our work is the first to use GPUs for parallelizing non-tail-recursive branching patterns in a load balanced manner.
We also reduce the memory usage of the intermediate graph representations by aggressively reducing the graph and inducing a subgraph before exploring the search tree.
This optimization and others enable launching more thread blocks to better utilize GPU resources, and using shared memory to quickly access intermediate graphs that are being used actively.
Our solution substantially outperforms the state-of-the-art GPU solution on a number of large and complex graphs.
\looseness=-1

\balance
\bibliographystyle{IEEEtran}
\bibliography{main}

\end{document}